\documentclass{ws-p8-50x6-00}

\usepackage{bm,float}

\begin{document}

\title{Particle-rotor model description of Doubly Odd T\lowercase {a} Isotopes}

\author{N. Itaco, L. Coraggio, A. Covello and A. Gargano}

\address{Dipartimento di Scienze Fisiche, Universit\`a di Napoli
Federico II \\
and Istituto Nazionale di Fisica Nucleare \\
 Complesso
Universitario di Monte S. Angelo, Via Cintia, 80126 Napoli, Italy}


\maketitle

\abstracts{
The doubly odd Ta isotopes are studied within the framework of the
particle-rotor model. The main aim of this study is to obtain information on
the neutron-proton interaction in deformed nuclei, with particular attention
focused on the role of the tensor force. To this end, both zero-range and
finite-range phenomenological interactions are considered. Comparison of the
calculated results with experimental data provides evidence of the
importance of the tensor-force effects. We have also performed calculations
using a realistic two-body $G$ matrix for the neutron-proton interaction. 
Some preliminary results are presented here.}

\section{Introduction}
The neutron-proton interaction plays an important role in the
description of doubly odd deformed nuclei. As is well known, the two
most important effects associated with the residual interaction
between the odd neutron and the odd proton are the
Gallagher-Moszkowski (GM) splitting\cite{Gallagher58} and the Newby
(N) shift.\cite{Newby62} Further information on the neutron-proton
interaction may be obtained by studying the odd-even
staggering in $K \ne 0$ bands.
In fact, this effect may be traced to direct Coriolis coupling of 
$K \ne 0$ bands with one or more
N-shifted $K = 0$ bands. Bands which exhibit odd-even 
staggering represent therefore a significant source of knowledge of 
the effective neutron-proton interaction.

In previous works,$^{3{\rm -}5}$ we have 
studied the doubly odd isotopes $^{174,176}$Lu
within the framework of the particle-rotor model. Our aim was to assess the
role of the effective neutron-proton interaction, with particular attention
focused on the tensor force. The results of these calculations showed that
the tensor force as well as the space-exchange and spin-spin space-exchange
force play an important role in the description of $K=0$ bands and of some $K
\not= 0$ bands.

Motivated by these findings, we have extended our study to other doubly odd
deformed nuclei in the rare-earth region. Here, we present some results
concerning the two heavier nuclei $^{180,182}$Ta. As we shall see in Sec.
3, the experimental data are very well reproduced when using the central
plus tensor force proposed long ago by Boisson {\it et
al.}\cite{Boisson76} 
This is in complete agreement with the results of our previous 
studies,$^{3{\rm -}5}$ confirming
that this effective neutron-proton interaction is quite suitable for the
unified model description of doubly odd rare-earth nuclei.  Being completely
phenomenological in nature, this interaction is of course in no way related
to the free nucleon-nucleon ($NN$) potential.

Over the past several years, realistic effective interactions derived from a
free $NN$ potential  have been successfully used in the shell-model
description of a number of near-magic nuclei.\cite{coraggio00,zako00} 
To our knowledge, no attempt
has instead been made to relate the effective neutron-proton interaction in
deformed nuclei to the free $NN$ interaction. 

In this situation, we have
found it challenging to turn our attention to this problem. To start with,
we have used as effective neutron-proton interaction the bare $G$ matrix
derived from the Bonn-A free $NN$ potential.\cite{machleidt87} 
In Sec. 4 we shall present some
results of these preliminary calculations which provide considerable
encouragement for further work in this direction.

\section{Outline of the Model and Calculations}
We assume that the unpaired neutron and proton are strongly coupled
to an axially symmetric core and interact through an effective 
interaction.
The total Hamiltonian is written as
\begin{equation}
H=H_0 + H_{{\rm RPC}} + H_{{\rm ppc}} + V_{np}.
\end{equation}
The term $H_0$ includes the rotational energy of the whole system, 
the deformed, axially 
symmetric field for the neutron and proton, and the intrinsic contribution
from the rotational degrees of freedom.
It reads
\begin{equation}
H_0 = \frac{\hbar^2}{2 {\cal J}} ( \bm{I}^2 - I_3^2 ) + H_n + H_p +
\frac{\hbar^2}{2 {\cal J}} [ ( \bm{j}_n^2 - j_{n3}^2 ) + 
( \bm {j}_p^2 - j_{p3}^2 ) ].
\end{equation}
The two terms $H_{{\rm RPC}}$ and $H_{{\rm ppc}}$ in Eq. (1) stand for 
the Coriolis coupling
and the coupling of particle degrees of freedom through the rotational motion,
respectively. Their explicit expressions are
\begin{equation}
H_{{\rm RPC}} = - \frac{\hbar^2}{2 {\cal J}} (I^+J^- + I^-J^+),
\end{equation}
\begin{equation}
H_{{\rm ppc}} = \frac{\hbar^2}{2 {\cal J}} (j_n^+j_p^- + j_n^-j_p^+).
\end{equation}
The effective neutron-proton interaction has the general form
\begin{equation}
V_{np}=V(r) [ u_0 + u_1 \bm{\sigma}_p \cdot \bm {\sigma}_n + 
u_2 P_M +u_3 P_M \bm {\sigma}_p \cdot \bm {\sigma}_n 
+ V_T S_{12} + V_{TM} P_M S_{12} ],
\end{equation}
with standard notation.\cite{Boisson76}
In our calculations we have used a finite-range force with a radial
dependence $V(r)$ of the Gaussian form 
\begin{equation}
V(r) = {\rm exp} (-r^2/r_0^2),
\end{equation}
as well as a zero-range force,
\begin{equation}
V_{np}^{\delta} = \delta (r) [ v_0 + v_1 \bm {\sigma}_p \cdot 
\bm {\sigma}_n ].
\end{equation}
As basis states we use the eigenfunctions of $H_0$, which, 
according to the assumed axial and reflection symmetry, take the form
\[
| \nu _n \Omega_n \nu_p \Omega_p I M K \rangle =  
\left ( \frac{2I + 1}{16 \pi^2} 
\right )^{\frac{1}{2}} [ D^I_{MK} | \nu _n \Omega_n \rangle | \nu_p \Omega_p
\rangle 
\]
\begin{equation}
~~~~~~~~~~~~~~~~~~~~~~~~~~ + (-)^{I+K} D^I_{M-K}  
|\nu_n \overline{\Omega}_n \rangle | \nu_p \overline{\Omega}_p \rangle ], 
\end{equation}
where the state $| \nu \overline{\Omega} \rangle$ is the time-reversal
partner of $| \nu \Omega \rangle$. 
We have used the standard Nilsson potential\cite{Gustafson67} to
generate the single-particle Hamiltonians $H_n$ and $H_p$. The
parameters $\mu$ and $\kappa $ have been fixed by using the mass-dependent
formulas of Ref. 11. The deformation
parameter $\beta_2$ has been deduced for each doubly odd isotope 
from the neighboring
even-even nucleus while the single-particle energies for the odd
proton and the odd neutron and the rotational parameter 
$\frac{\hbar^2}{2 {\cal J}} $ have been derived from the experimental
spectra of the two neighboring odd-mass nuclei.

For the neutron-proton interaction we have used both a finite-range
force with a Gaussian radial shape and a $\delta$ interaction. As
regards the former, we have explored the role of the tensor
force by performing two different calculations, with and without
the tensor terms. In both cases for the parameters of the interaction
we have adopted the values determined by Boisson {\it et al.}
\cite{Boisson76} in their study of doubly odd nuclei in the
rare-earth region. As regards the $\delta$ force, we have used 
for the strength of the
spin-spin term $v_1$ the value -0.20 MeV, which leads to the
lowest possible disagreement between theory and experiment. More
details about the choice of the parameters of the potentials as well
as their explicit values can be found in Ref. 3.

We shall give an outline of the calculations performed with the $G$
matrix derived from the Bonn-A potential in Sec. 4.

\section{Results and Comparison with Experiment}

In this section we present some selected results of our study of
the two isotopes $^{182}$Ta and $^{180}$Ta. More precisely, we consider the
lowest $K^{\pi} = 0^-$ band in $^{182}$Ta and the $K^{\pi} = 1^+$
ground-state band in $^{180}$Ta. A more complete presentation and
discussion of our results will be given in a forthcoming paper.

In Fig. 1 we compare the experimental spectrum\cite{nndc} of the 
$K^{\pi} = 0^-$ $p \frac{7}{2} [404] n \frac{7}{2} [503]$ band in 
$^{182}$Ta with the spectra obtained by using the 
Gaussian force, with and without tensor terms, and the $\delta$ force.
We see that the right level order and an excellent agreement with
experiment is obtained when using the Gaussian force with tensor
terms. The discrepancy between the calculated and experimental excitation
energies is at most about 20 keV.
This is not the case for the calculations performed with the pure
central finite-range interaction and the $\delta$ force. In fact, as
shown in Fig. 1, these calculations do not reproduce the right level
order. This may be traced back to the different values of the
calculated N-shift. They are 28, 1, and -6 keV in case (a), (b), and (c), 
respectively, to be compared with the experimental value of 26 keV.
\begin{figure}[H]
\begin{center}
\epsfxsize=100mm 
\epsfbox{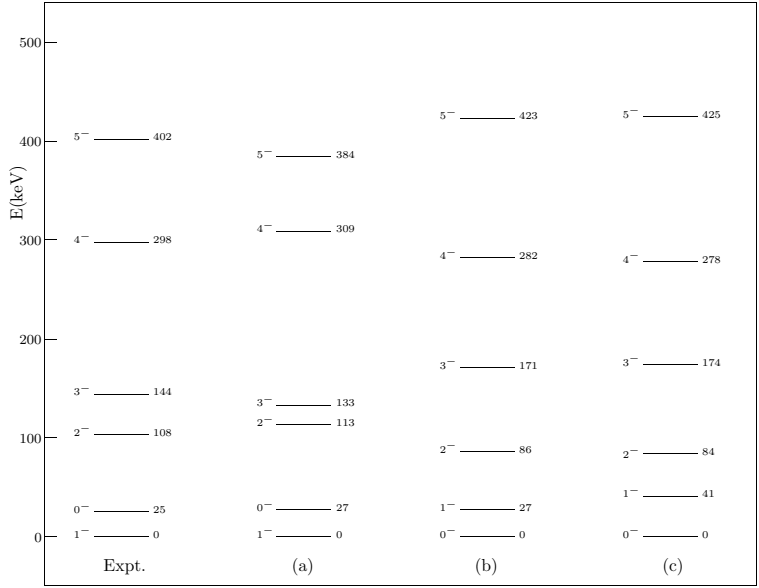} 
\end{center}
\caption{Experimental and calculated spectra of the lowest $K^{\pi} =
0^-$ band in $^{182}$Ta. The theoretical spectra have been obtained by
using (a) a central plus tensor force with a Gaussian radial shape,
(b) a Gaussian central force, and (c) a $\delta$ force.} 
\end{figure}
\noindent
Therefore, while in case (a) the calculated N-shift is in very good 
agreement with experiment, it becomes too small in case (b), and in
case (c) it has even the wrong sign.  

Let us now come to the $K^{\pi} = 1^+ p \frac{7}{2} [404] n \frac{9}{2} [624] $
ground-state band in $^{180}$Ta. This band exhibits a rather large odd-even
staggering, as can be seen from Fig. 2 where the
experimental\cite{Dracoulis98,Saitoh99} ratio
$[E(I)-E(I-1)]/2I$ is plotted vs $I$ and compared with
the calculated ones.
\begin{figure}[H]
\begin{center}
\epsfbox{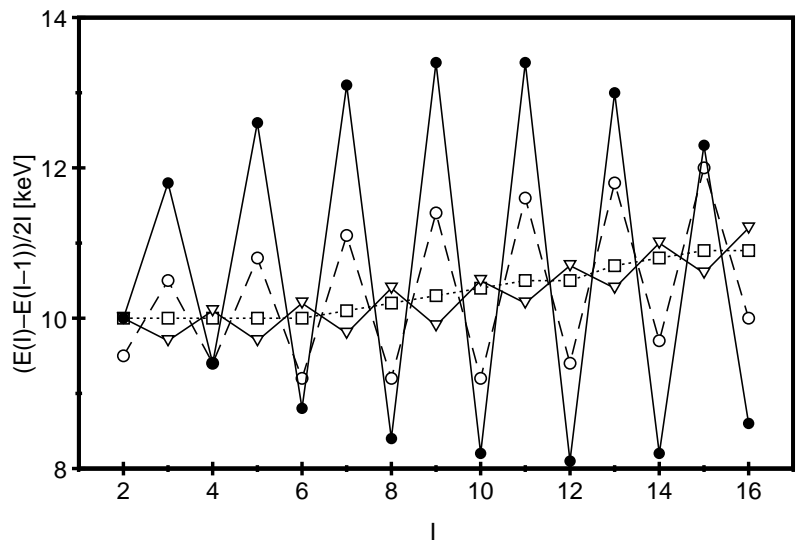} 
\end{center}
\caption{Experimental and calculated odd-even staggering of the
$K^{\pi} = 1^+$ ground-state band in $^{180}$Ta. Solid circles correspond to
experimental data. The theoretical results are represented by open
circles (Gaussian central plus tensor force), squares (Gaussian
central force), and triangles ($\delta$ force).} 
\end{figure}
\vspace{-2mm}
We see that the experimental behavior is well reproduced by
the calculation including the tensor force. When using the pure
central Gaussian force the staggering is almost nonexistent and in the
case of the $\delta$ force not only its magnitude is very small, but
it has also the opposite phase.
It should be pointed out that the staggering in this band may be
traced to direct Coriolis coupling with the $K^{\pi} = 0^+ p
\frac{7}{2} [404] n \frac{7}{2} [633] $. The fact that only the
calculation including the tensor terms gives the right staggering
implies therefore that only this force is able to produce a sizeable N shift
for this  $K^{\pi} = 0^+$ band. Since this band has not been
recognized in $^{180}$Ta, a direct comparison is not possible at
present.
It should be noted, however, that the lowest $K^{\pi} = 0^+$ band
observed in $^{174}$Lu corresponds just to the same intrinsic n-p 
configuration.
In our study\cite{sagata} of $^{174}$Lu it turned out that only the 
Gaussian plus tensor
force is able to produce a spectrum of this band in good agreement
with the experimental one.

\section{Calculations with a Realistic Two-Body $G$ Matrix}
As already mentioned in the Introduction, we present here some results of
preliminary calculations performed by using as neutron-proton residual
interaction the bare $G$ matrix derived from the Bonn-A free $NN$ 
potential.\cite{machleidt87}
The $G$ matrix is defined\cite{krenc76} by the integral equation
\begin{equation}
G(\omega)=V+V Q_2 {{1} \over{\omega-Q_2TQ_2}} Q_2G(\omega), 
\end{equation}
where $V$ represents the $NN$ potential, $T$ denotes the two-nucleon kinetic 
energy, and $\omega$ is the so-called starting energy. 
The operator $Q_{2}$ is the Pauli exclusion operator for two
interacting
nucleons, and its complement $P_2=1-Q_2$ defines the space within which the
$G$ matrix is calculated. All the states outside the $P_2$ space are
intermediate states represented by plane wave functions. In spherical
shell-model calculations the $P_2$ space is defined in terms of harmonic
oscillator eigenvectors. For deformed nuclei, we have calculated the $G$
matrix making use of Nilsson basis states.
The calculation of the $G$ matrix is performed in an essentially exact way
by using the Tsai-Kuo method.\cite{tsai72}

In Tables 1 and 2 we report the calculated and experimental GM splittings
for $^{182}$Ta and $^{180}$Ta.
We see that while all the signs are correctly reproduced 

\begin{table}[H]
\begin{center}
\caption{Experimental and calculated GM splittings (keV) in $^{182}$Ta.}
\begin{tabular}{|cccccc|}
\hline
\multicolumn{2}{|c} {Configuration} & & & & \\
Proton & Neutron & $K_<^\pi$ & $K_>^\pi$ & Expt. & Calc.\\
\hline
 & & & & &\\
$7/2[404]$ & $1/2[510]$ & $3^-$ & $4^-$ & -100 & -172 \\
$7/2[404]$ & $3/2[512]$ & $2^-$ & $5^-$ & 139 & 374 \\
$7/2[404]$ & $7/2[503]$ & $0^-$ & $7^-$ & -121 & -146 \\
$9/2[514]$ & $1/2[510]$ & $4^+$ & $5^+$ & 148 & 149 \\
$9/2[514]$ & $3/2[512]$ & $3^+$ & $6^+$ & -98 & -105 \\
$5/2[402]$ & $1/2[510]$ & $2^-$ & $3^-$ & 114 & 84 \\
 & & & & & \\
\hline
\end{tabular}
\end{center}
\end{table}

\noindent
also the
quantitative agreement  can be considered satisfactory, the largest
discrepancy with experiment being 100 keV in one case only. As regards the N
shift, we obtain 46 keV for the the
lowest $K^{\pi} = 0^-$ band in $^{182}$Ta, to be compared with the experimental
value of 26 keV.

\begin{table}[H]
\begin{center}
\caption{Same as Table 1, but for $^{180}$Ta.}
\begin{tabular}{|cccccc|}
\hline
\multicolumn{2}{|c} {Configuration} & & & & \\
Proton & Neutron & $K_<^\pi$ & $K_>^\pi$ & Expt. & Calc.\\
\hline
 & & & & &\\
$7/2[404]$ & $9/2[624]$ & $1^+$ & $8^+$ & -100 & -186 \\
$7/2[404]$ & $5/2[512]$ & $1^-$ & $6^-$ & -88 & -78 \\
 & & & & & \\
\hline
\end{tabular}
\end{center}
\end{table}

\section{Summary}
In this paper, we have presented some results of a particle-rotor model
study of doubly odd Ta isotopes aimed at obtaining detailed information on
the effective neutron-proton interaction in the rare-earth region. In
particular, we have explored the role of tensor force, which has been
neglected in most of the existing studies to date. A comparison of our
results with experimental data evidences the importance of this force for
the description of $K=0$ bands and of some $K \not=0$ bands, thus confirming
the results of our previous study of Lu 
isotopes.$^{3{\rm -}5}$

As regards the results obtained by using a realistic two-body $G$-matrix
interaction, which does not contain any adjustable parameter, 
they are certainly better than
what one would expect. This shows that effective interactions derived from
the free $NN$ potential may be suitable for use in the unified model
description of deformed nuclei. Further work in this direction is in
progress.

\section*{Acknowledgments}
This work was supported in part by the Italian Ministero
dell'Universit\`a e della Ricerca Scientifica e Tecnologica (MURST). 
NI thanks the European Social Fund for financial support.

\end{document}